\documentclass[useAMS,usenatbib]{mn2e}
\usepackage{epsfig}
\usepackage{times}

\def\msun{\mbox{$\textrm M_\odot$}}

\def\ga{\mathrel{\hbox{\rlap{\hbox{\lower4pt\hbox{$\sim$}}}{\raise2pt\hbox{$>$}}}}}
\def\la{\mathrel{\hbox{\rlap{\hbox{\lower4pt\hbox{$\sim$}}}{\raise2pt\hbox{$<$}}}}}
\def\sdss1212{\mbox{SDSS\,J$121209.31+013627.7$}}
\def\S05{Schmidt et~al. 2005}
\def\mdot{\mathrel{\mbox{\.{M}\,}}}

\begin{document}
\title[SDSS\,J$121209.31+013627.7$]{The nature of the 
close magnetic white dwarf $+$ probable brown dwarf
  binary SDSS\,J121209.31+013627.7\thanks{Based on
    observations made with the Faulkes Telescope North, the Isaac
    Newton Telescope and the {\it Wide Field Camera}, the William Herschel
    Telescope and {\it ULTRACAM} high speed photometer, and the {\it Swift} 
space observatory.}}

\author[M. R. Burleigh et~al.]
{M. R. Burleigh$^1$
 T. R. Marsh$^2$  B. T. G\"ansicke$^2$ M. R. Goad$^1$
V. S. Dhillon$^3$ S. P. Littlefair$^3$ \newauthor 
M. Wells$^4$ N. P. Bannister$^1$ 
C. P. Hurkett$^1$ A. Martindale$^1$ 
P. D. Dobbie$^1$
S. L. Casewell$^1$ \newauthor
D. E. A. Baker$^1$ J. Duke$^1$ 
J. Farihi$^5$ M. J. Irwin$^6$ P. C. Hewett$^6$
P. Roche$^7$ F. Lewis$^7$
  \\
$^1$ Department of Physics and Astronomy, University of Leicester,
Leicester LE1 7RH, UK \\
$^2$ Department of Physics, University of Warwick, Coventry CV4 7AL,
UK\\
$^3$ Department of Physics and Astronomy, University of Sheffield,
Sheffield S3 7RH, UK\\
$^4$ Oundle School, Oundle, Northamptonshire, UK \\
$^5$ Gemini Observatory, Hilo, Hawaii, USA \\
$^6$ Institute of Astronomy, University of Cambridge, Cambridge CB3
0HA, UK\\
$^7$ Department of Physics and Astronomy, University of Wales, 
Cardiff CF24 3YB, UK\\
}

\date{Accepted 2100 December 32. Received 2099 December 25; 
in original form 1888 October 11}

\pagerange{\pageref{firstpage}--\pageref{lastpage}} \pubyear{2006}

\maketitle

\label{firstpage}

\begin{abstract}

Optical time series photometry of the short period magnetic white dwarf $+$
probable brown dwarf binary \sdss1212 reveals pulse-like 
variability in all bands from $i'$ to $u'$, increasing towards bluer
wavelengths and peaking at $u'$. These
modulations are most likely due to a self-eclipsing accretion hot spot on the
white dwarf,
rotating into view every 88.43~minutes. This period is commensurate
with the H$\alpha$ radial velocity period determined by \citet{sdss1212} of
$\approx90$~minutes, and consistent with the rotation period of the
accretor being equal to the binary orbital period. 
We combine our observations with other recently reported 
results to provide an accurate ephemeris.  
We also detect the system in X-rays with {\it Swift}, and estimate the
accretion rate at $\approx10^{-13}$\msun~yr$^{-1}$. 
We suggest that \sdss1212 is most likely a magnetic cataclysmic variable
in an extended state of very low accretion, similar to the
well-studied polar EF~Eri. 
Alternatively, the putative brown dwarf is not filling its Roche Lobe
and the system is a detached binary in which the white dwarf is
efficiently accreting from the wind of the secondary. 
However, it is unclear whether an L~dwarf wind is strong
enough to provide the measured accretion rate. We suggest further
observations to distinguish between the Roche Lobe overflow and wind
accretion scenarios.

\end{abstract}


\section{Introduction}

Brown dwarf companions to white dwarfs are rare \citep*{fbz05}. 
Proper motion surveys and searches for infrared (IR) excesses have so far
found only three confirmed examples: GD\,165 (DA$+$L4,
\citealt{becklin88}),  
GD\,1400 (DA$+$L$6/7$, \citealt{GD1400}; 
\citealt{dobbie05}), and WD\,0137$-$349 (DA$+$L8, \citealt{maxted06}; 
\citealt{wd0137b}). 
GD\,165 is a widely separated system (120\,AU) and the separation
of the components in GD\,1400 is currently unknown, but WD\,0137$-$349
is a close ($P=116$~minutes), detached (non-interacting) binary. The
L8 companion must have survived a phase of Common Envelope (CE)
evolution during which the orbit decayed to near the current period. 

There is no
confirmed brown dwarf secondary in a cataclysmic variable 
(CV). Many are strongly
suspected to contain such very low mass companions, but the
secondaries are difficult
to certify \citep*{lvm}. 
For example, the claimed direct detection of the L\,$4/5$ secondary
in the magnetic CV (polar) EF\,Eri \citep{howell01} has subsequently been
called into question through evidence for IR cyclotron emission 
\citep{harrison04}. Recent radial velocity measurements also support a
sub-stellar mass for the secondary in EF~Eri, although the unknown 
white dwarf mass still allows for a stellar companion \citep{Howell06}. 

\sdss1212 (hereafter SDSS\,1212) was identified by \citet{sdss1212}
from the Sloan Digitised Sky Survey as a $g'\approx18$, 
$T_{\rm eff}\approx10,000$~K, 
$B\approx13$~MG magnetic DA white dwarf. Optical spectra display the 
characteristic 
Zeeman split H$\alpha$ and H$\beta$ absorption lines and also a 
weak, narrow H$\alpha$ 6563{\AA} emission line, which is variable both in 
radial velocity and flux. This emission line is interpreted as 
arising from the irradiated face of a cool secondary star in a close orbit 
with the white dwarf. The radial velocity variations show a $\approx90$~minute 
periodicity, which is most likely the orbital period of the system.     
Despite a photometric measurement at 1.2\,microns (J band), the
cool secondary was not detected by \citet{sdss1212}, limiting its absolute
magnitude to M$_{\rm J} >13.4$, equivalent to a brown dwarf of spectral
type L\,5 or later ($T_{\rm eff} < 1700$K). 

The lack of evidence for
ongoing accretion led \citet{sdss1212} to conclude that SDSS\,1212 is
most likely a detached binary. 
As with WD\,0137$-$349, the cool 
companion would always have been a brown dwarf and, therefore, 
must have survived a previous phase of  
CE evolution. In this picture, SDSS\,1212 would be a 
precursor of a magnetic CV (a pre-polar), and the first known  
magnetic white dwarf with a cool secondary \citep{liebert05}. 
Alternatively, \citet{sdss1212} suggested that SDSS\,1212 is a
magnetic CV in an extended low state, similar to that experienced by
EF~Eri between 1996 and early 2006. 

SDSS\,1212 is clearly an important new binary requiring more detailed study. 
Therefore, we obtained time series optical photometry of 
the system to search for variability to: (a) better determine 
the orbital period; (b) investigate the effects of irradiation
on the putative brown dwarf; (c) to search for a possible short ($<5$~minute)
eclipse of the white dwarf, which would yield a radius for the cool
companion; and (d) to search for non-radial pulsations, since the
temperature of the white dwarf is close to the range occupied by the
ZZ\,Ceti stars. Our surprising 
results prompted us to then obtain X-ray 
observations with the {\it Swift} space observatory, which provide further
insight into the nature of the system.

\section{Observations and data reduction}

\subsection{Faulkes Telescope North and Isaac Newton Telescope}

 

SDSS\,1212 was remotely 
observed on 2006 January 3 (MJD\,53738) for one hour with the 2m robotic 
Faulkes Telescope North (FTN), 
located at the Haleakala Observatory on Maui, Hawaii.  
The observations were made through a
Bessell~$V$ filter with 45~second exposures 
obtained approximately every 2~minutes.  
The debiassed and flat-fielded data were downloaded and differential 
aperture photometry was performed with respect to three 
comparison stars in the $4.6' \times 4.6'$ 
field of view, using the Starlink software package {\it PHOTOM}. 
Once we had checked the comparison stars for variability, they
were added together and divided into the target to give the
differential photometry. The resultant light 
curve is shown in Figure~1 (top). The target is clearly variable at the  
$\approx3\%$ level, consistent with the 
$\approx90$~minute radial velocity period reported by \citet{sdss1212}.

\begin{figure}
\caption{Top: One hour, $V$-band light curve of SDSS\,1212 obtained 
with the 2m robotic Faulkes Telescope North on 2006 January
  3 (MJD\,53738), revealing variability at the $\approx3\%$
  level. Bottom: Three hour, 
$r'$ light curve of SDSS\,1212 obtained with the INT WFC on
  2006 February 4 (MJD\,53771). The time axes have been scaled to
  match for ease of comparison.}
\label{fig:ftintdata}
\psfig{file=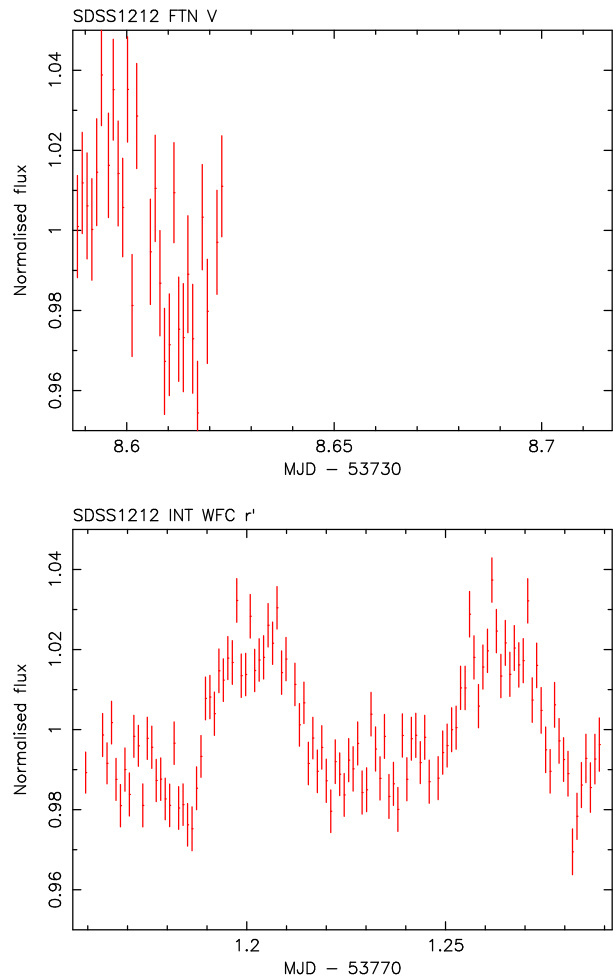,angle=0,width=8.cm}

\end{figure}


Although the FTN data revealed for the first time 
that SDSS\,1212 is photometrically
variable, the data did not cover the entire radial velocity
period and were of relatively poor quality. Therefore, 
two of us (MRB and SLC) obtained further  
time series photometry with the {\it Wide Field Camera} (WFC)
on the 2.5m Isaac Newton Telescope (INT) in La Palma  
on 2006 February 4 (MJD\,53771). The target was observed 
continuously for 3~hours through a Sloan $r'$ filter, with 60~second 
exposures. The readout speed meant that exposures were 
obtained every $\approx1.5$~minutes. 
The weather was clear and photometric throughout the observations, 
with seeing $\approx1''$. 
The data were reduced with the WFC pipeline  
reduction software at the Cambridge Astronomical Survey 
Unit, Institute of 
Astronomy, Cambridge. Differential photometry was performed in the same 
manner as the FTN data, although four different non-variable comparison stars 
were utilised. The light curve is also shown in Figure~1 (bottom). 



\subsection{{\it ULTRACAM} on the William Herschel Telescope}

SDSS\,1212 was independently observed by three of us (TRM, VSD, SPL)  
on the nights of 2006 March 11 and 12 (MJD\,53805 \& MJD\,53806) 
using the multi-band high-speed CCD camera {\it ULTRACAM} \citep{ultracam}  
mounted on the 4.2m William Herschel Telescope (WHT) in La
Palma. We observed simultaneously 
in $i'$, $g'$ and $u'$ with exposure times of 10s (with only
25~milli-seconds of dead time betweem each frame) for
a total of 3.6~hours on the first night (during gaps in the main programme) and
for 2~hours on the second night. The weather was clear throughout the
observations, but the seeing which was $\approx1''$ on the first night,
deteriorated to $2-3''$ on the second night. The seeing combined with a bright
Moon meant that the first night's data were superior.
The data were reduced with the {\it ULTRACAM} data reduction pipeline software. 
The reduced, normalized differential light curves from the first night
are shown in
Figure~2. Note that the 10s exposures have been binned to
90s in this plot. 

\begin{figure}
\caption{Simultaneous $i'$, $g'$ and $u'$ light curves of SDSS\,1212
  obtained with {\it ULTRACAM} on the WHT on 2006 March 11 (MJD\,53805). 
The original 10s exposures have
  been binned to 90s per data point. The $g'$ and $u'$ datasets have been
  arbitrarily offset for clarity. The fitted model consists of
  parabolas plus three Gaussians; the fit to the $g'$ data yields a
  period of 88.43~minutes. }
\label{fig:whtdata}
\psfig{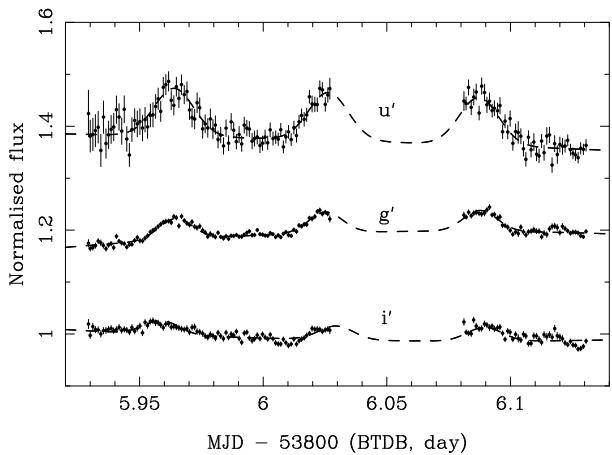}
\end{figure}

\subsection{{\it Swift} observations}

The primary scientific objective of the 
{\it Swift} space observatory is to detect and obtain follow-up 
multi-wavelength observations of Gamma-Ray Bursts (GRBs). In addition
to a gamma ray detector (Burst Alert Telescope, BAT), {\it Swift}
carries an X-ray telescope (XRT) and an ultraviolet and optical
telescope (UVOT). During periods when GRBs and their afterglows
are not being observed, these instruments can be used to observe other
interesting astronomical targets. 
Following our discovery of the unusual optical
light curves of SDSS\,1212 (discussed in detail below), we requested
and were awarded such ``fill-in'' time 
with {\it Swift} to search for  
X-ray emission. These observations were carried out in late
2006 April and May (Table 1). 
SDSS\,1212 was observed on 11 separate occasions for $\approx1$~ksec per
observation.  

\subsubsection{{\it Swift} XRT data reduction}

The {\it Swift} XRT PC mode event lists were
processed using the standard XRT data reduction software {\it xrtpipeline} 
version 0.9.4, within FTOOLS v6.0.3, screening for hot and flickering pixels,
bad columns, and selecting event grades 0-12, giving a total on-source
exposure time of $10.7$~ksec.

An X-ray source is clearly detected 
at $8\sigma$ close to the position of SDSS\,1212, at 
$12$hr~$12$m~$09.7$s, $+01$d~$36$m~$25.8$s (J2000), with an error radius
of $5.2''$ ($90\%$ containment), and has a mean X-ray count rate of
$2.6\pm0.6\times10^{-3}$~ct~s$^{-1}$ in the $0.3 - 10$~keV range. 
We note that no
X-ray source was detected at this position in the {\it ROSAT} all sky
survey. 

\begin{table}
\caption{Journal of {\it Swift} observations.}
\label{tab:swifttimes}
\begin{center}
\begin{tabular}{cccc}
\hline
Obs. ID & Start Time & Start Time & Exp. \\
      & (UT) & (MJD) & (sec.)\\
\hline
\hline
00030391001 &   2006-04-20 01:01:01  & 53845.0423727 &  794.4 \\
00030391002 &   2006-04-23 01:41:00  & 53848.0701389 &  956.9\\
00030391003 &   2006-04-25 00:18:00  & 53850.0125000 & 794.3\\
00030391004 &   2006-04-27 21:14:00  & 53852.8847222 & 985.9\\
00030391005 &   2006-04-30 10:27:01  & 53855.4354282 & 966.8\\
00030391006 &   2006-05-02 12:23:30  & 53857.5163194 & 531.8\\
00030391007 &   2006-05-04 16:56:01  & 53859.7055671 & 797.3\\
00030391009 &   2006-05-11 21:26:00  & 53866.8930556 & 810.5\\
00030391011 &   2006-05-16 13:52:01  & 53871.5777893 & 872.1\\
00030391012 &   2006-05-18 04:18:01  & 53873.1791782 & 1237.5\\
00030391013 &   2006-05-20 06:13:01  & 53875.2590394 & 2030.3\\
\hline
\end{tabular}
\end{center}
\end{table}


\section{Data analysis and results}



\subsection{Optical light curves}

Although the FTN $V$~band data from 2006 January revealed the 
photometric variability of SDSS\,1212, the pulse-like behaviour is 
not due to
the variations in flux of the weak H$\alpha$ emission line reported
by \citet{sdss1212}. The INT WFC and WHT {\it ULTRACAM} 
multi-band datasets (Figures~1, 2 \& 3) 
reveal that the variability is larger in the 
ultraviolet (UV) and blue
compared to the red, and that in all bands the light curve consists of
a broad Gaussian peak rising from an underlying steady continuum
level. The source of the variability is most likely a hot spot on the
surface of the white dwarf, which is self eclipsed every time the star
rotates. This type of variability is commonly observed in polars 
(e.g. \citealt{stockman}; \citealt*{gaensicke95}).  

To determine the rotation period of the white dwarf from this hot
spot, we fit the {\it ULTRACAM} $g'$ band light
curve from 2006 March with a model consisting of a  
parabola (allowing for any general trend during the observations, 
due to colour-dependent extinction effects)  
plus 3~Gaussians (2~for the second night), 
which are constrained to have the same width
and height but are otherwise free. 
We then extended our model to
include the 2006 February INT WFC data, and to include the 
very recently reported 
data of \citet{koen}, who observed SDSS\,1212 on several occations in
white light, $V$ and $R$ filters 
for a total 27 hours between 
2006 March 23 (MJD\,53823) and 2006 April 26 (MJD\,53851). 

The pulse times for all these observations 
are given in Table~2.  
All times are barycentric corrected. 
We find no ambiguity in the cycle
identifications, and from a linear fit to the pulse times 
we determine the following ephemeris: 

\vspace{2mm}
\noindent
MJD (BTDB) $=  T_0 + P E = 53798.5953(2) + 0.0614081(7) E$\\

The zeropoint $T_0$ was chosen to minimise
the correlation coefficient between it and the period $P$.
The errors on the last two digits are given in parenthesis and 
are $1\sigma$ uncertainties. $E$ is the number of cycles since $T_0$
\footnote{By extrapolating back to the 2006 January FTN observation,
  we can use this ephemeris to verify that the peak in those data is at
  phase~0 (cycle number $-977$).}. 
The rotation period of $88.428 \pm0.001$~minutes compares favourably with 
the $93\pm2$~minutes H$\alpha$ radial velocity period estimated by 
\citet{sdss1212}, the $88\pm1$~minutes
$K_s$-band variability recently 
reported by \citet{debes} (presumably due to
cyclotron emission) and the similar 88.43~minutes  
optical photometric period reported by \citet{koen}.
Since the H$\alpha$ emission line originates on the secondary star,
while the photometric variability is due to a hot spot on the white
dwarf, we conclude that 
the binary is synchronously rotating, at least 
within the large uncertainty on the spectroscopic period, and has 
most likely evolved to this state through magnetic locking. 
 
\begin{table}
\caption{Pulse times from model fits to the 2006 March {\it ULTRACAM} 
observations, the 2006 February INT observations and the recent
results of \citet{koen}.}
\label{tab:pulsetimes}
\begin{center}
\begin{tabular}{cccc}
\hline
Dataset & Cycle & Time & Uncertainty\\
     & ($E$) & MJD (BTDB) & 1$\sigma$\\
\hline
\hline
INT/WFC $r'$ & -446 & 53771.20735 & 0.00051\\
INT/WFC $r'$ & -445 & 53771.26823 & 0.00052\\
WHT/{\it ULTRACAM} $g'$ & 120  &    53805.96444 & 0.00032\\
WHT/{\it ULTRACAM} $g'$ & 121  &    53806.02519 & 0.00051\\
WHT/{\it ULTRACAM} $g'$ & 122  &    53806.08779 & 0.00041\\
WHT/{\it ULTRACAM} $g'$ & 137 &    53807.00771 & 0.00103\\
Koen \& Maxted & 398 &  53823.0375 & 0.0025\\
Koen \& Maxted & 412 &  53823.8946 & 0.0025\\
Koen \& Maxted & 413 &  53823.9567 & 0.0025\\
Koen \& Maxted & 414 &  53824.0188 & 0.0025\\
Koen \& Maxted & 430 &  53825.0000 & 0.0025\\
Koen \& Maxted & 446 &  53825.9842 & 0.0025\\
Koen \& Maxted & 447 &  53826.0452 & 0.0025\\
Koen \& Maxted & 462 &  53826.9621 & 0.0025\\
Koen \& Maxted & 851 &  53850.8522 & 0.0025\\
Koen \& Maxted & 852 & 53850.9101 & 0.0025\\
\hline
\end{tabular}
\end{center}
\end{table}

We also fit the simultaneous $i'$, $g'$ and $u'$ band light curves
obtained by {\it ULTRACAM} on 2006 March 11-12 
(Figure~2) with the same
model to determine the pulse heights in each case. We find the height
of the pulse is $2.6\% \pm 0.1\%$ at $i'$, $4.0\% \pm
0.1\%$ at $g'$, and $9.2\% \pm 0.4\%$ at $u'$.
\footnote{Note that \citet{sdss1212} give the magnitudes of SDSS\,1212 in
these bands as $i' = 18.24\pm0.02$, $g' = 17.99\pm0.02$ and  
$u' = 18.43\pm0.03$.} The formal uncertainties on 
these fits may underestimate the systematic uncertainties, particularly at 
$i'$. 

Assuming the pulses
originate solely from a hot spot on the white dwarf, we can model
the {\it ULTRACAM} $u'g'i'$ light curves of SDSS\,1212 and estimate the 
temperature of the spot (Figure~3). 
We used the code described by
\citet{gaensicke98} and \citet{gaensicke06}, and find the spot has a
temperature $T_{\rm spot}\simeq14,000$~K. The self-eclipse of the spot
constrains its size, although the inclination is somewhat of a guess; 
the large radial velocity amplitude measured by \citet{sdss1212}
suggests it cannot be particularly low, and the system is not eclipsing
so it cannot be higher than $\approx75^\circ$. 
We estimate the fractional area of the spot is
$\approx5\%$ of the white dwarf surface, in good agreement with the values
found in other systems (e.g.~A\,M~Her, \citealt{gaensicke06}). 
We also estimate the luminosity of the hot
polar cap at $\approx1.6\times10^{29}$~erg~s$^{-1}$. 

While an eyeball fit of a nonmagnetic model spectrum to the SDSS\,1212
spectrum suggests 
$T_{\rm eff} \approx11,000$~K, a fit to the lower envelope of the
{\it ULTRACAM} light curves gives $T_{\rm eff} \approx9,500$~K (Table~3).  
For this temperature, and 
assuming a radius of $8 \times 10^8$~cm for a $\approx0.6\msun$ white dwarf, 
the distance is $\approx120$~pc.

\begin{figure}
\caption{Model fits to the (from top to bottom) 
{\it ULTRACAM} $u'g'i'$ light curves to derive
  the parameters of the hot spot, as
  described by \citet{gaensicke98} and \citet{gaensicke06}. The
  parameters are given in Table~3.}
\label{fig:spotfits}
\psfig{file=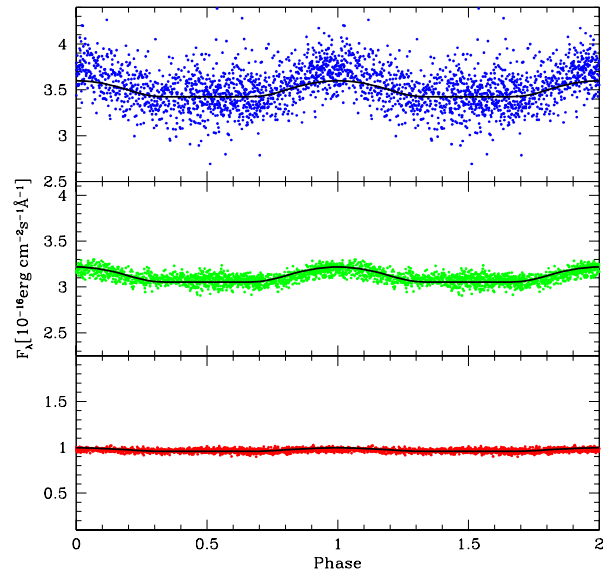,angle=0,width=8.cm}
\end{figure}

\begin{table}
\caption{Hot spot parameters derived from model fits to the {\it ULTRACAM} 
$u'g'i'$ light curves, for a white dwarf of radius 
$8\times10^8$~cm.}
\label{tab:spotfit}
\begin{center}
\begin{tabular}{cc}
\hline
Parameter & Value\\
\hline
\hline
$T_{\rm WD}$ & $9,500$~K\\
Distance & $120$~pc\\
$T_{\rm spot}$ & $14,000$~K\\
Spot opening angle & $50^{\circ}$\\
Colatitude of spot & $75^{\circ}$\\
\hline
\end{tabular}
\end{center}
\end{table}

\citet{sdss1212} report that the H$\alpha$ emission line disappears at one 
phase,  and conclude that the system is likely viewed at high
inclination. However, we find no evidence for eclipses.
The white dwarf also lies near the edge of the 
ZZ~Ceti instability strip for (non-magnetic) H-rich degenerates, but  
we see no non-radial pulsations in our optical 
photometry. This suggests its temperature is most likely 
below the lower edge of the instability strip, assuming that the  
boundaries are the same for magnetic and non-magnetic white dwarfs.

We also note that \citep{koen} believed that they had found a
secondary pulse between the maxima in the optical light
curves, possibly caused by a second pole. 
Although there is some slight evidence for such features in
Figures 1 \& 2, our data are inadequate to support such a
claim. 

\subsection{{\it Swift} XRT observations}


The XRT count rate is too low to detect any X-ray variability on the
orbital period, but sufficient 
to construct an X-ray spectrum (Figure~4). 

\begin{figure}
\caption{{\it Swift} XRT X-ray spectrum of SDSS\,1212. The fit is for 
the single temperature thermal plasma model described in the text.}
\label{fig:spotfits}
\psfig{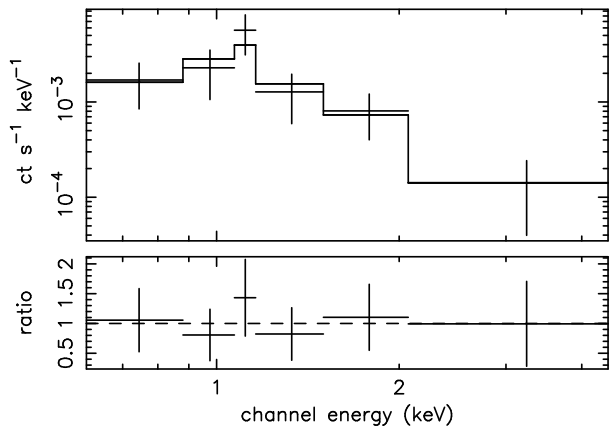}
\end{figure}

For the spectral extraction, we specified a 20~pixel radius circular aperture
centred on the X-ray position. For the background, we chose an annular region
of inner radius 30~pixels, outer radius 50~pixels, centred on the same
position. Source and background spectra were extracted from the cleaned event
lists using {\it Xspec11}, using the appropriate ancilliary response
files and response matrices as defined in the latest release of
the {\it Swift} CALDB (version 24). 
Due to the small number of source counts, the XRT spectrum was grouped
to a minimum of 5 
counts per bin and model fits were minimised using Cash statistics (C-stat).

We attempted a number of model fits to the extracted spectrum; 
an absorbed blackbody, an absorbed bremsstrahlung and an absorbed one 
temperature thermal plasma (the {\it MEKAL} model in {\it Xspec}). 
Since the X-ray
count rate and signal/noise are low, we initially 
fixed the Hydrogen (H) column density in each model at
a number of values between $10^{18}$~atoms~cm$^{-2}$ and 
$10^{20}$~atoms~cm$^{-2}$. This is a reasonable assumption 
for the distance and
Galactic latitude and longitude of SDSS\,1212 (\citealt{Paresce};  
\citealt{Frisch}). 

The blackbody fit gives a temperature $kT_{\rm BB}=0.36$~keV for 
$N_{\rm H} = 10^{18}$~atoms~cm$^{-2}$ (C-stat$=4.1$ for 4
degrees of freedom), and $kT_{\rm
  BB}=0.35$~keV for $N_{\rm H} = 10^{20}$~atoms~cm$^{-2}$ (C-stat$=4.6$). 
The unabsorbed flux for these models is 
$\approx9 \times 10^{-14}$~erg~cm$^{-2}$~s$^{-1}$ (between
$0.3-10$~keV in all these calculations). 


Perhaps a more physically plausible model is a
bremsstrahlung spectrum, again with excess Hydrogen column
density. Free-free (bremsstrahlung) radiation is expected to dominate
the cooling of the gas in the post-shock region 
where accreting gas impacts the white dwarf
atmosphere. For a fixed column density 
$N_{\rm H}=10^{18}$~atoms~cm$^{-2}$,
the temperature $kT_{\rm Br}=1.64$~keV (C-stat$=4.6$ for 4~degrees of
freedom). The fit improves slightly as the H column density
increases. In fact, if the H column density is included as a free
parameter in the fit, we find that the best
fit bremsstrahlung model has a temperature $kT_{\rm Br}=0.92$~keV, 
and $N_{\rm H}=2\times10^{21}$~atoms~cm$^{-2}$ (C-stat$=3.7$ for
3~degrees of freedom).  For this
model the absorbed and unabsorbed flux is $8.9 \times 10^{-14}$ and $2.0 
\times 10^{-13}$~erg~cm$^{-2}$~s$^{-1}$. 


The best fit to the data is given by the single-temperature thermal
plasma model. We find $kT = 1.9$~keV (C-stat$=0.94$ for 4~degrees of
freedom), and the fit is very insensitive
to the assumed H column density. 
In particular, this model matches the flux 
at 1.1~keV, which is under-predicted in the blackbody and
bremsstrahlung models. This model fit is shown in Figure~4.  
The unabsorbed flux given by the 
single-temperature thermal plasma model is $1.2  
\times 10^{-13}$~erg~cm$^{-2}$~s$^{-1}$.  

Using the unabsorbed flux from the best-fit
single-temperature thermal plasma 
model, and assuming a distance of $120$~pc,  
we then estimate the X-ray luminosity $L_x = 2 \times 10^{29}$~erg~s$^{-1}$. 
Assuming this X-ray luminosity is entirely due to accretion, we can
estimate the accretion rate from:  

\begin{equation}
L_{\rm x} = \frac {GM_{\rm WD}\mdot\,} {R_{\rm WD}}
\label{eq:accretionrate}
\end{equation}

Assuming 
a canonical mass for the white dwarf $M_{\rm WD} = 0.6\msun$,  
$R_{\rm WD} = 8 \times 10^8$~cm and d$=120$~pc, 
we obtain an accretion rate of $\mdot\,
\approx3\times10^{-14}$\msun~yr$^{-1}$. For a distance of 
$145$~pc \citep{sdss1212}, 
$L_x = 3 \times 10^{29}$~erg~s$^{-1}$ and 
$\mdot\, \approx5\times10^{-14}$\msun~yr$^{-1}$.

Given the X-ray accretion rate estimated here, and the optical/UV
luminosity estimated in Section~3.1, the total accretion rate for 
SDSS\,1212 $\mdot_{\rm total}\, \approx10^{-13}$\msun~yr$^{-1}$. 

\section{Discussion}

Multi-band optical photometry of the magnetic white dwarf $+$ probable 
brown dwarf close binary SDSS\,1212 reveals a hot spot on the surface of the 
white dwarf, which rotates into view every 88.4~minutes. This is in addition 
to the weak, narrow, and 
variable (in radial velocity and strength) H$\alpha$ 
emission line observed by \citet{sdss1212} in optical spectra, which most likely arises 
from the warm irradiated face of the probable brown dwarf secondary. If the 
system is synchronously rotating, 
then the orbital period is the same as the 88.43~minute rotation 
period of the white dwarf. 

The large hot spot on the white dwarf is directly linked to magnetically
funneled accretion. A likely hypothesis for the heating of the white dwarf
surface in this spot is irradiation with cyclotron radiation and/or
thermal bremsstrahlung \citep{gaensicke95}, although remnant heat 
from a previous high state \citep{gaensicke06}, or heating by direct
accretion can not be excluded. 
This hypothesis is confirmed by the detection of the system 
in X-rays. A $\approx10,000$~K white dwarf photosphere does not have 
any emission in this waveband. These results 
strongly suggest that SDSS\,1212 is in fact a magnetic CV 
in a low accretion state, in which the brown dwarf
secondary fills its Roche Lobe during periods of active accretion.  
We note that although \citet{sdss1212} concluded that SDSS\,1212 was
in their view
more likely a detached binary, they also suggested that it could be a
low accretion rate polar.  
The optical light curves presented here are  
similar to those seen in other polars in low states, e.g. EF~Eri
\citep{harrison04}, 
further strengthening the polar identification. 

The comparison with EF~Eri is appropriate for a number of
reasons. That system was first identified as an active X-ray source in
1978 and appeared in a high state every time it was observed either in
the optical or X-rays until it almost entirely switched off in
1996. It only re-appeared as an actively accreting CV early in
2006. This decade-long low state shows that polars can efficiently
hide from discovery. \citet{beuermann00} modelled the optical low
state spectrum of EF~Eri entirely in terms of emission from the white
dwarf with a temperature $T_{\rm eff}=9,500$~K at a distance of
$\approx130$~pc. Based on the non-detection of TiO absorption bands in the
I-band, the authors suggested the spectral type of the donor star to
be later than L1. \citet{howell01} later suggested they had directly
detected the secondary in a near-IR spectrum, but
\citet{harrison04} showed that, for EF~Eri at least, cyclotron emission due
to low level accretion spoils the chances of a direct detection in
this waveband.

The detection of a heated polar cap on SDSS\,1212 lends strong support
to the polar scenario. 
The first evidence for a relatively large moderately heated spot on
white dwarfs 
 in polars was derived by \citet{hv88} for
the prototype system AM~Herculis on the basis of {\it IUE} 
ultra-violet spectra. 
This hot spot
has been quantitatively modelled by \citet{gaensicke95} 
in terms of a heated cap near the
magnetic pole of the white dwarf. The luminosity of the polar cap in 
AM~Her could be explained either by heating through ongoing accretion, or
as a remnant from deep localised heating of the envelope during a
previous high state. The best-documented evidence for a heated spot
on AM~Her during the low state is based on {\it HST} / {\it FUSE} 
data \citep{gaensicke06}.
Evidence for heated polar caps has been 
found in low state observations of many systems: 
V834~Cen shows a
strong modulation in the U-band \citep{ferrario92}, 
DP~Leo in the UV \citep{stockman}. Spectroscopic signatures of  heated
polar caps have been found in V1043~Cen (RX~J$1313.2-3259$, 
\citealt{gaensicke00}), and in V834~Cen, MR~Ser, and BL~Hyi
\citep{araujo}. 

From modelling the observed UV 
cyclotron emission of the high-field polar AR\,UMa in
low state, \citet{gaensicke01} derived an accretion rate of $\mdot\, \approx2
\times 10^{-13}\msun$~yr$^{-1}$. For the same system, \citet{Szkody99} 
determined a low state X-ray luminosity of $\approx2 \times
10^{29}$~erg~s$^{-1}$. These measurements are comparable to our
estimates of the X-ray luminosity and accretion rate of SDSS\,1212,
supporting the low state, magnetic CV scenario. 

\citet{debes} recently detected cyclotron emission in the $K_s$~band,
also indicating that the white dwarf is accreting material 
from its very low mass companion. By comparing their ephemeris with
our optical ephemeris, we note that the cyclotron hump is co-incident
with the optical peak, as expected if the emission is from the same
spot. While it is unclear whether cyclotron emission is responsible
for all the near-IR excess reported by 
\citet{debes}, they use the $H$ and $K_s$ band  
minimum fluxes to place an upper limit of L7 on the secondary 
spectral type.

The 88.43~minutes orbital period, well outside the CV minimum period of
80-82~minutes, 
combined with the evidence for a very
low mass secondary, and the cool temperature of the white dwarf primary, 
suggests that SDSS\,1212 could be a candidate
for a period bouncer, i.e. a CV that has evolved through and beyond the
period minimum. 
These highly evolved binaries are
predicted by population models to be the most abundant species of CVs
\citep{kolb}, but have so far been elusive despite intensive
searches.
\citet*{patterson05} present evidence for a number of such systems among
non-magnetic dwarf novae. 
In this scenario, the probable brown dwarf 
secondary may have evolved to that state through mass loss, 
and could plausibly have 
originally evolved through the CE phase as 
a higher mass main sequence M~dwarf. 





An alternative possibility is that the secondary does not fill its
Roche Lobe, and accretion is occuring via efficient capture of almost
all of the stellar wind \citep*{Li}. 
\citet{Schmidt05b} discuss the evolutionary status of six low accretion
rate ($\mdot\, \la 10^{-13}$\msun\,yr$^{-1}$) binaries  
with cool ($T_{\rm eff}<14,000$K) magnetic ($B \approx 60$MG)
white dwarf primaries, termed Low Accretion Rate
Polars (LARPs) by \citet{Schwope}. 
The first, HS\,1023$+$3900, was discovered by \citet{Reimers99}. 
These binary periods are all $\ga2.5$~hr, and \citet{Schmidt05b} suggest 
that the secondaries are too late in spectral type to fill their Roche Lobes. 
If they are not in contact, these binaries can be considered as
post-CE, pre-polar systems. Such binaries are also of interest
because the accretion of the stellar wind by the primary could lead to
diminution of the wind-driven angular momentum loss which usually
drives the period evolution of these systems. 
However, all of the claimed pre-polar binaries were
discovered through optical cyclotron emission features, which
lie in the IR for SDSS\,1212. 
All the claimed pre-polars are also faint in X-rays 
($L_{\rm x} \la 10^{29}$~erg~s$^{-1}$, \citealt{Szkody04}), 
and none contain a substellar secondary. If SDSS\,1212
is a wind-accreting pre-polar, then it would be the first to be
discovered without optical cyclotron lines. 

The wind accretion scenario requires a strong brown dwarf wind. 
\citet{ww06} suggest that coronal M~dwarf X-ray luminosities 
$\approx10^{29}$~erg~s$^{-1}$ can drive several
$10^{-13}\msun$~yr$^{-1}$ of material, which is then siphoned over
and accreted by the magnetic white dwarf in LARPs. 
Young ($\la0.1$~Gyr) brown dwarfs, which have spectral types M5-M9,
have been detected in X-rays with luminosities $L_{\rm x} \approx
10^{27}$~erg~s$^{-1}$ (\citealt{Stelzer04}; \citealt{Preibisch}). 
However, no
L~dwarf has been detected in X-rays, limiting their 
quiescent X-ray luminosities
to $L_{\rm x} \la 10^{23}$~erg~s$^{-1}$ \citep{Stelzer06}.  
This large suppression in X-ray luminosity between young and old brown
dwarfs is almost certainly due to the transition to a largely neutral
atmosphere \citep{Stelzer06}. It is likely that the rotation period
of the brown dwarf in SDSS\,1212 is the same as the orbital period,
i.e.~88.4~minutes. Field brown dwarfs typically rotate with periods of
order $2 - 20$~hours \citep{Bailer-Jones}. Thus, the brown dwarf in 
SDSS\,1212 might have been spun up by a factor~10 during and after the
CE phase. Adopting the rotation-activity relation for field dwarfs
allows us to place upper limits on the likely X-ray flux, assuming the
activity is not saturated. A spin up of factor 10 gives a two orders
of magnitude increase in X-ray luminosity \citep{Pizzolato}. The
largest X-ray flux we might then expect from the companion is $L_{\rm x}
\la 10^{25}$~erg~s$^{-1}$, too small by a factor $\sim10^4$ to drive
the material being accreted by the white dwarf. 

For similar reasons we are not persuaded that the X-ray flux detected
by {\it Swift} is due to intrinsic activity on the brown dwarf. Even
if the L-dwarf was an unlikely source of at least some of 
the X-ray emission seen by {\it
  Swift}, the detection of cyclotron emission in the near-IR by
\citet{debes} proves that accretion is taking place. However, 
\citet{Rottler} point out that the H$\alpha$ emission from the K~dwarf
in the pre-CV binary V471~Tau 
disappeared between 1985 and 1992. Therefore, it
cannot be due to irradiation. They also point out that the
temperatures of the white dwarfs in many pre-CV binaries do not
seem high enough to account for irradiation-induced H$\alpha$
emission. A combination of irradiation and
chromospheric activity may be needed to account for the 
H$\alpha$ emission lines from the secondaries in these systems. The 
$T_{\rm eff}\approx10,000$K white dwarf in SDSS\,1212 may also not be hot
enough to produce the H$\alpha$ emission seen by \citet{sdss1212}
through irradiation, and
the L~dwarf may be intrinsically active (but not necessarily a
detectable X-ray source). 
Indeed, \citet{Howell06} claim that the radial velocity
  variable H$\alpha$ emission line observed from the putative
  sub-stellar companion in EF~Eri is due to intrinsic activity rather
  than irradiation.


Despite the Sloan Digitised Sky Survey more than doubling
the numbers of known white dwarfs with detached cool companions 
\citep{Silvestri} and
the numbers of magnetic white dwarfs (e.g.~\citealt{Vanlandingham}), 
\citet{liebert05} noted that there was zero overlap
between these two samples. In contrast, an estimated 
$25\%$ of accreting CVs have a magnetic white dwarf primary
\citep{WF}. 
If SDSS\,1212 is a polar in a low state,
then the lack of magnetic white dwarfs in detached binaries
remains. \citet{liebert05} discuss a number of explanations, and
suggest that the presence of the companion and the likely large mass
and small radius of the magnetic white dwarf 
(relative to nonmagnetic degenerate dwarfs) may provide a selection 
effect against the discovery of such binary systems.

More observations are needed before the nature of SDSS\,1212 can be
fully confirmed. Detailed modelling of the magnetic 
white dwarf's optical spectrum will better determine its temperature, 
surface gravity, mass, radius, field strength and configuration and 
cooling age. 
Further optical spectroscopy is necessary to obtain a better
radial velocity curve and place the  spectroscopy and photometry onto
a single ephemeris. Near- and mid- IR spectroscopy will lead to
a detailed model of the cyclotron emission and may help to
reveal the spectral type of the secondary. 
A direct detection of the brown dwarf 
would also allow investigation of the effects of
irradiation. Finally, UV  light curves may
help to distinguish between the low-state polar and wind accretion
models, since different accretion spot geometries
might be expected in each case. Recent {\it GALEX} observations of
EF~Eri reveal a large-amplitude modulation in the far- and near-UV,
which \citet{Szkody06} attempted to fit with a white dwarf plus heated
polar cap. While their simplistic model using blackbodies rather than
model atmopsheres remains inconclusive, it is clear that UV
observations of SDSS\,1212 have the potential to constrain the hot
spot temperature and geometry.  

\section{Acknowledgments}

MRB \& BTG acknowledge the support of PPARC Advanced Fellowships. 
TRM acknowledges the support of a PPARC Senior Fellowship. SPL \& PDD 
thank the support of PPARC PDRAs. CPH, SLC \& AM   
acknowledge the support of PPARC Postgraduate Studentships. 
This paper has made use of observations obtained with the Faulkes
Telescope North on Maui, Hawaii, and we thank the help of 
students at Oundle School, Northamptonshire, UK in obtaining the
data. The Isaac Newton and William Herschel Telescopes are operated on the 
island of 
La Palma by the Isaac Newton Group in the Spanish Observatorio del Roque de los
Muchachos of the Instituto de Astrofisica de Canarias. 
{\it ULTRACAM} is supported by PPARC grants PP/D002370/1 and
PPA/G/S/2003/00058. We thank the
Director of the {\it Swift} mission for the rapid award of fill-in time
to make the X-ray observations. We thank Richard Jameson, Andrew King 
and Graham Wynn for useful discussions on the nature of SDSS\,1212.  

\bibliographystyle{mn}
\bibliography{mbu}

\label{lastpage}

\end{document}